\documentclass[twocolumn,prl,aps,superscriptaddress,nobalancelastpage]{revtex4}
\usepackage{graphicx,bm}
\begin{document}

\title{Evolution of the Fermi surface of BaFe$_2$(As$_{1-x}$P$_x$)$_2$ \\
on entering the superconducting dome}

\author{H. Shishido}
\affiliation{Department of Physics, Kyoto University, Sakyo-ku, Kyoto 606-8502, Japan.}
\affiliation{Research Center
for Low Temperature and Materials Sciences, Kyoto University, Sakyo-ku, Kyoto 606-8501, Japan.}
\author{A.F. Bangura}
\affiliation{H. H. Wills Physics Laboratory, Bristol University, Tyndall Avenue, BS8 1TL, United Kingdom.}
\author{A.I. Coldea}
\affiliation{H. H. Wills Physics Laboratory, Bristol University, Tyndall Avenue, BS8 1TL, United Kingdom.}
\author{S. Tonegawa}
\affiliation{Department of Physics, Kyoto University, Sakyo-ku, Kyoto 606-8502, Japan.}
\author{K. Hashimoto}
\affiliation{Department of Physics, Kyoto University, Sakyo-ku, Kyoto 606-8502, Japan.}
\author{S. Kasahara}
\affiliation{Research Center for Low Temperature and Materials Sciences, Kyoto University, Sakyo-ku, Kyoto 606-8501,
Japan.}
\author{P.M.C. Rourke}
\affiliation{H. H. Wills Physics Laboratory, Bristol University, Tyndall Avenue, BS8 1TL, United Kingdom.}
\author{H. Ikeda}
\affiliation{Department of Physics, Kyoto University, Sakyo-ku, Kyoto 606-8502, Japan.}
\author{T. Terashima}
\affiliation{Research Center for Low Temperature and Materials Sciences, Kyoto University, Sakyo-ku, Kyoto 606-8501,
Japan.}
\author{R. Settai}
\affiliation{Graduate School of Science, Osaka University, Toyonaka, Osaka 560-0043, Japan.}
\author{Y. \={O}nuki}
\affiliation{Graduate School of Science, Osaka University, Toyonaka, Osaka 560-0043, Japan.}
\author{D. Vignolles}
\affiliation{Laboratoire National des Champs Magn\'{e}tiques Intenses (CNRS), Toulouse, France.}
\author{C.~Proust}
\affiliation{Laboratoire National des Champs Magn\'{e}tiques Intenses (CNRS), Toulouse, France.}
\author{B. Vignolle}
\affiliation{Laboratoire National des Champs Magn\'{e}tiques Intenses (CNRS), Toulouse, France.}
\author{A. McCollam}
 \affiliation{High Field Magnet Laboratory, Institute for Molecules and Materials, Radboud University Nijmegen, Toernooiveld 7, 6525
ED Nijmegen, The Netherlands}
\author{Y.~Matsuda}
\affiliation{Department of Physics, Kyoto University, Sakyo-ku, Kyoto 606-8502, Japan.}
\author{T.~Shibauchi}
\affiliation{Department of Physics, Kyoto University, Sakyo-ku, Kyoto 606-8502, Japan.}
\author{A. Carrington}
\affiliation{H. H. Wills Physics Laboratory, Bristol University, Tyndall Avenue, BS8 1TL, United Kingdom.}

\begin{abstract}
Using the de Haas-van Alphen effect we have measured the evolution of the Fermi surface of
BaFe$_2$(As$_{1-x}$P$_x$)$_2$ as function of isoelectric substitution (As/P) for $0.41<x<1$ ($T_c$ up to 25\,K). We
find that the volume of electron and hole Fermi surfaces shrink linearly with decreasing $x$. This shrinking is
accompanied by a strong increase in the quasiparticle effective mass as $x$ is tuned toward the maximum $T_c$. It is
likely that these trends originate from the many-body interaction which gives rise to superconductivity, rather than
the underlying one-electron bandstructure.
\end{abstract}

\maketitle

Superconductivity in the 122 iron-pnictide family XFe$_2$As$_2$ (where X=Ba,Sr or Ca), can be induced by a variety of
means, including doping \cite{rotter08,sefat08}, pressure \cite{alireza} or isoelectric substitution either on the Fe
\cite{paulraj_superconductivity_2009} or As sites \cite{jiang_xu}. The highest $T_c$ achievable by each of these routes
is roughly the same. It has been suggested \cite{MazinSJD08,KurokiUOAA09,Chubukov2008} that the interband coupling
between the hole and electron sheets plays an important role in determining the magnetic or superconducting order
formed at low temperature.  Discovering how the Fermi surface evolves as a function of the various material parameters
which drive the material from an antiferromagnetic spin density wave state, through the superconducting dome and
eventually towards a paramagnetic non-superconducting metal, should therefore be an important step toward gaining a
complete understanding of the mechanism that drives high temperature superconductivity in these materials.

In the antiferromagnetic state is it expected that the Fermi surface suffers a major reconstruction. This is supported
experimentally by the observation \cite{analytisba122,harrison,sebastian} of very low frequency quantum oscillations in
the undoped XFe$_2$As$_2$ compounds, corresponding to very small Fermi surface pockets which are 1-2\% of the total
Brillouin zone planar area.  At the other extreme of the phase diagram, where the materials are paramagnetic and
non-superconducting, quantum oscillation measurements of SrFe$_2$P$_2$ \cite{analytis09} and CaFe$_2$P$_2$
\cite{coldea09} show that the Fermi surface is in good agreement with conventional bandstructure calculations. Up to
now, tracking the changes in the Fermi surface across the phase diagram using quantum oscillations has not been
possible because of the additional disorder and high  $H_{c2}$ introduced by doping. Measurements on the low $T_c$
(6\,K) superconducting iron-pnictide LaFePO \cite{coldea08,sugawara} established that the Fermi surface is in broad
agreement with bandstructure with moderate correlation enhancements of the effective mass. It is not clear whether the
higher $T_c$ pnictide superconductors, which unlike LaFePO occur in close proximity to a magnetic phase, are also well
described by bandstructure and whether the electronic correlations change significantly.

\begin{figure*}
\centering
\includegraphics[width=16cm]{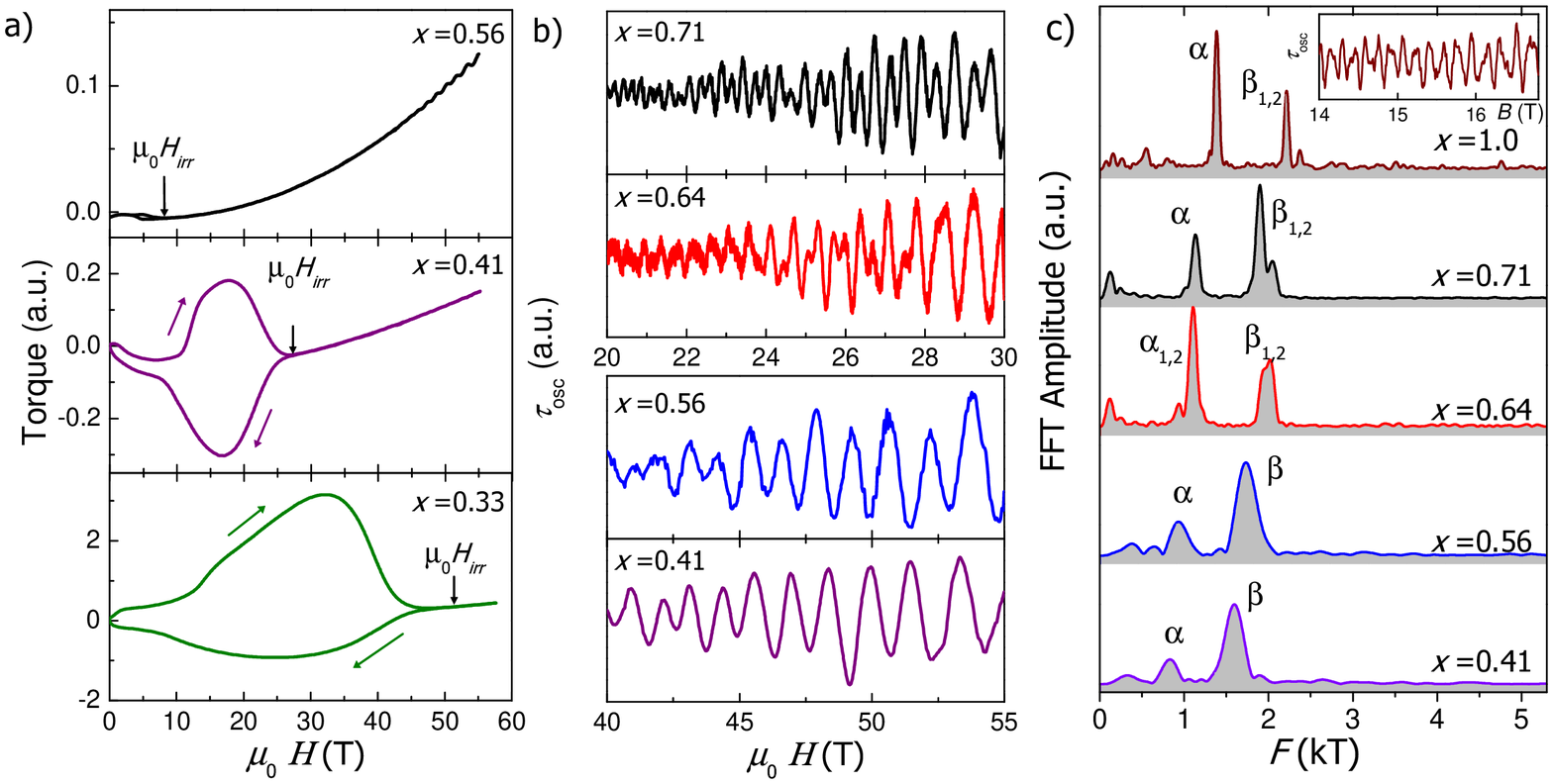}
\caption{(color online). a) Raw torque data for BaFe$_2$(As$_{1-x}$P$_x$)$_2$ with $x=0.33, 0.41, 0.56$ measured at
1.5~K. b) The oscillatory part of torque for selected samples and c) the corresponding Fourier transform for $x=0.41,
0.56$ between 35-55~T at 1.5~K and $x=0.64, 0.71$ between 20-30~T at 0.35~K as well as $x=1$ between 12-16.8~T at
0.092~K. The inset shows the oscillatory signal for $x=1.0$.}
 \label{Fig:RawData}
\end{figure*}

The substitution of P for As in the series BaFe$_2$(As$_{1-x}$P$_x$)$_2$ offers an elegant way to suppress magnetism
and induce superconductivity without doping \cite{Kasahara0905.4427}. As P and As are isoelectric, there is no net
change in the ratio of electrons to holes and the system remains a compensated metal for different $x$ (equal volumes
for the electron and hole Fermi surfaces). This series has several remarkable properties which are similar to those
observed in cuprate superconductors. Firstly, the temperature dependence of the resistivity changes from a quadratic
($\rho\sim T^2$) to linear behavior ($\rho\sim T$) as the system evolves from a conventional Fermi liquid ($x=1$)
towards the maximum $T_c$ ($x=0.33$). Secondly, there is strong evidence from magnetic penetration depth
\cite{Hashimoto0907.4399}, thermal conductivity \cite{Hashimoto0907.4399} and NMR \cite{Nakai09} measurements that for
$x=0.33$ the superconducting gap has line nodes.

In this Letter, we report the observation of quantum oscillation signals in samples of BaFe$_2$(As$_{1-x}$P$_x$)$_2$ as
$x$ is varied across the superconducting dome from $x=1$ to $x=0.41$ with $T_c \sim 25$~K ($\simeq 0.8\,T_{c}^{\rm
max}$). Our data show that the Fermi surface shrinks and the quasiparticle masses become heavier as the material is
tuned toward the magnetic order phase boundary, at which $T_{c}$ reaches its maximum. This implies that the significant
change of electronic structure caused by many-body interactions plays a key role for the occurrence of unconventional
high-$T_{c}$ superconductivity.

Single crystal samples of BaFe$_2$(As$_{1-x}$P$_x$)$_2$ were grown as described in Ref.\ \cite{Kasahara0905.4427}.  The
$x$ values were determined by an energy dispersive X-ray analyzer. Measurements of the quantum oscillations in magnetic
torque [the de Haas-van Alphen (dHvA) effect] were made using a miniature piezo-resistive cantilever technique.
Experiments were performed in: a dilution refrigerator system with DC fields up to 17\,T (Osaka), a pumped $^3$He
system with DC fields up to 30\,T (Nijmegen) and 45\,T (Tallahassee) and a pumped $^4$He system with pulsed fields up
to 55\,T (Toulouse). Our data are compared to band structure calculations which were performed using the
\textsc{wien2k} package \cite{wien2k}.

Fig.~\ref{Fig:RawData}(a) shows raw magnetotorque $\tau(H)$ data for three values of $x$ measured up to 55\,T at
$T\simeq$\,1.5\,K, with the magnetic field direction close to the $c$-axis. The torque response for the higher $T_c$
samples ($x=0.33$, $T_c\simeq 30$\,K and $x=0.41$, $T_c\simeq 25$\,K) is highly hysteretic. $\mu_0H_{irr}$ increased
substantially with $x$ reaching a maximum of 51.5\,T for the highest $T_c$ sample ($x=0.33$), which is close to the
estimated $H_{c2}$ value \cite{Hashimoto0907.4399}. By subtracting a smooth polynomial background the oscillatory dHvA
signal is clearly seen above $H_{irr}$ (Fig.~\ref{Fig:RawData}(b)) for all samples except for that with the highest
$T_c$ ($x=0.33$).

\begin{figure*}[t]
\centering
\includegraphics[width=16cm]{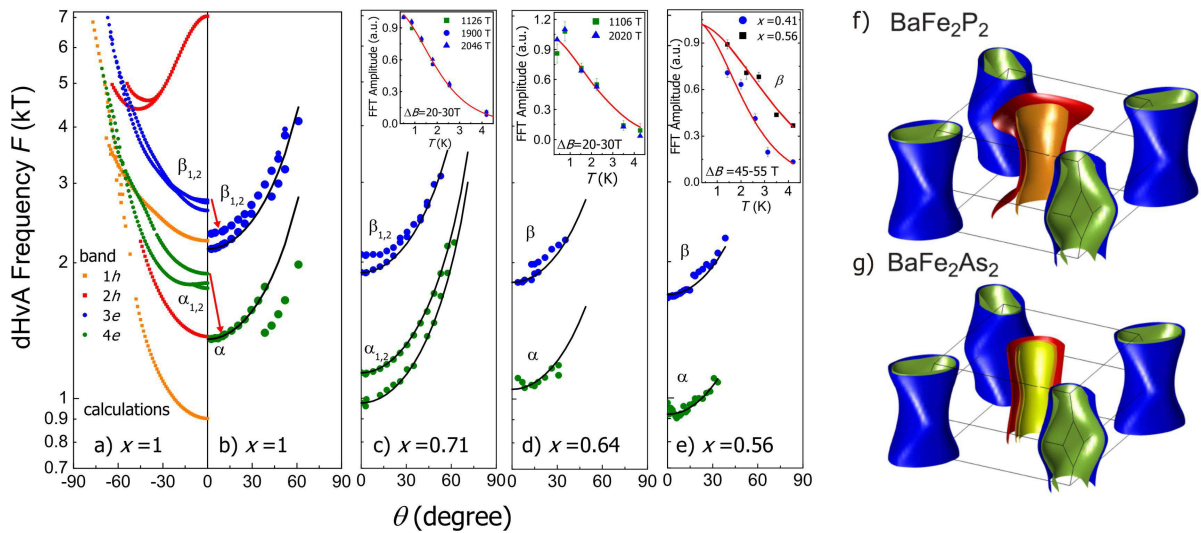}
\caption{(color online). a) Angle dependence of the predicted orbits from
 $B\parallel$[001] ($\theta=0^{\circ}$) to $B\parallel$[100]($\theta=90^{\circ}$) for $x=1$ (BaFe$_2$P$_2$)
  and the electron orbits observed experimentally for b) $x=1$, c) $x=0.72$, d) $x=0.64$, e) $x=0.56$.
Solid lines correspond to $F(\theta=0)/\cos \theta$. The insets show the temperature dependence of the Fourier
amplitudes for each composition, which are fitted (solid line) to the Lifshitz-Kosevich formula $X/\sinh X, X=14.69
\,m^*T/B$ to determine the effective masses, $m^*$ \cite{shoenberg}. The calculated Fermi surfaces of the end members,
f) BaFe$_2$As$_2$ (non-magnetic) and g) BaFe$_2$P$_2$ are also shown.} \label{Fig:Ftheta}
\end{figure*}

From the fast Fourier transform (FFT) spectra of the oscillatory data (Fig.~\ref{Fig:RawData}(c)) we can extract the
dHvA frequencies $F$. These are related to the extremal cross-sectional areas, $A_{k}$, of the Fermi surface orbits
giving rise to the oscillations via the Onsager relation, $F = (\hbar/2\pi e) A_{k}$. From the evolution of these
frequencies (see Fig.\ \ref{Fig:Ftheta}) as the magnetic field is rotated from being along the $c$ axis ($\theta =
0^{\circ}$) towards being perpendicular the $c$ axis ($\theta = 90^{\circ}$), we can deduce the shape of the Fermi
surface.

Band structure calculations of the Fermi surface of the end members of the BaFe$_2$(As$_{1-x}$P$_x$)$_2$ series are
shown in Fig.\ \ref{Fig:Ftheta}. For both BaFe$_2$As$_2$ and BaFe$_2$P$_2$ the calculations were done with the
experimental lattice parameters and pnictide $z$ position in the non-magnetic state. The two electron sheets at the
zone corner are quite similar in size and shape in both compounds but there are significant differences between the
hole sheets. In BaFe$_2$As$_2$ three concentric quasi-two-dimensional hole tubes are located at the zone center,
whereas in BaFe$_2$P$_2$ the inner one of these tubes is absent whereas the outer tube has become extremely warped. The
Fermi surface of  BaFe$_2$P$_2$ is quite similar to that for SrFe$_2$P$_2$ \cite{analytis09}.

For SrFe$_2$P$_2$\cite{analytis09}, CaFe$_2$P$_2$ \cite{coldea09} and LaFePO \cite{coldea08} it was observed that the
strongest dHvA signals came from the electron sheets and so this is likely also to be the case for BaFe$_2$P$_2$. Thus,
we assign the two strongest peaks ($\alpha$ and $\beta$) to the inner and outer electron sheet, respectively (see Fig.\
\ref{Fig:RawData}(c)). The observed frequencies for $x=1$ are somewhat smaller than those predicted by the calculations
and a rigid band shift of $45-75$\,meV is needed to bring the orbits into agreement with experiment. This is quite
comparable to the shifts $50-60$\,meV needed for the electron sheets in SrFe$_2$P$_2$ \cite{analytis09}.  Although
there are other weak peaks in the FFT which could come from the hole sheets, further measurements with better signal to
noise ratio are needed to confirm these.

\begin{figure}
\centering
\includegraphics[width=7cm]{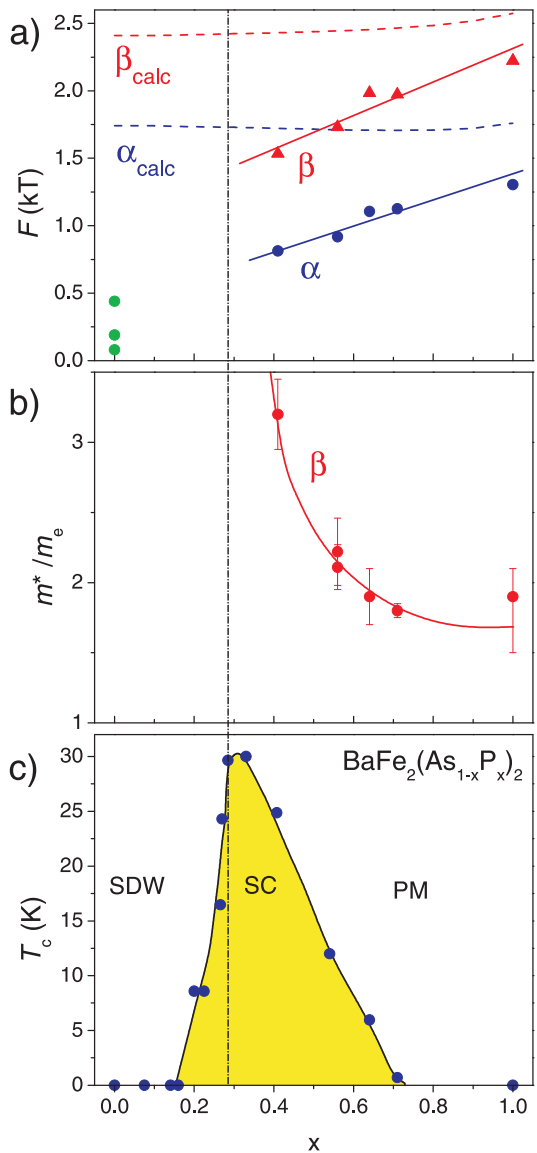}
\caption{color online. (a) Experimental (solid symbols) average electron sheet frequencies ($\alpha$ and $\beta$)
versus P content, $x$. The data for $x=0$ are taken from Ref.\ \cite{analytisba122}. The dashed lines show
bandstructure predictions. b) The variation with $x$ of the quasiparticle effective masses, $m^*$ and (c) $T_c$ after
Ref.\ \cite{Kasahara0905.4427}. The vertical dashed line marks the location of the onset of the appearance of magnetism
at $T=0$.} \label{Fig:AllvsX}
\end{figure}

As the sample composition is varied towards optimal doping, the signal from the electron sheets is reduced but two
peaks are clearly visible over the full doping range in sufficiently high fields (see Fig.\ \ref{Fig:RawData}(c)); the
mean free path for the $\beta$ orbits decreases from $\ell \sim 800$\AA~ to 170~\AA~ when $x$ varies from $x=1$ to
0.41. Importantly, we observe that the frequency of both these electron orbits decreases linearly with decreasing $x$
(see Fig.\ \ref{Fig:AllvsX}(a)). For $x=1$ the $\beta$ frequency equals $2.30$\,kT  which decreases to $1.55$\,kT for
$x=0.41$. Although we do not clearly observe them, charge neutrality means that the hole sheets must shrink by the same
factor.

The effective masses of the quasiparticles on the various orbits were determined by fitting the temperature dependent
amplitude of the oscillations to the Lifshitz-Kosevich formula \cite{shoenberg} (see insets of Fig.\ \ref{Fig:Ftheta}).
As shown in Fig.\ \ref{Fig:AllvsX}(b), the effective mass increases significantly as $x$ approaches the spin density
wave ordered phase. At the same time $T_c$ increases and shows a maximum at the boundary of the magnetic order.  As no
such increase is expected from the bandstructure (see later) the result implies a significant rise in the strength of
the many-body interactions, most likely caused by spin-fluctuations. A similar striking rise in the mass enhancement
close to an antiferromagnetic quantum critical point has also been observed in quasi-two dimensional heavy-Fermion
systems \cite{ShishidoSHO05}.

For $x=1$ bare masses calculated from the bandstructure for the electron sheets vary from 0.8 to 1.0 $m_e$. To obtain a
dHvA frequency of 1.55\,kT, appropriate for $x=0.41$, requires a rigid shift of the energy of band 4 (see Fig.\
\ref{Fig:Ftheta}(a)) by 170\,meV.  The bare mass then decreases from 0.92 to 0.81 $m_e$, in sharp contrast to the large
rise seen in experiment. The many body enhancement factor $m^*/m_b$ therefore increases from $\sim 2$ to $\sim 4$ as
$x$ decreases from 1 to 0.41. Note that conventional electron-phonon coupling is weak in these materials, and is
estimated to only account for $\sim $25\% of the observed mass enhancement \cite{Yildirim2009}.

The shrinking of the electron and hole sheets as $x$ decreases could come from either the underlying one-electron
bandstructure or alternatively it could be driven by many-body correlation effects. We have estimated the change in the
electron sheet area by calculating the bandstructure for both BaFe$_2$As$_2$ and BaFe$_2$P$_2$ with values of the
lattice constants and $z$ fixed to the experimental values appropriate for the various values of $x$. Experimentally it
is found that these parameters follow Vegard's law and scale linearly with $x$ \cite{Kasahara0905.4427}.  We find that
for both the As and P material the $\alpha$ and $\beta$ frequencies vary relatively little with $x$. In Fig.\
\ref{Fig:AllvsX}(a) we show a weighted average of these calculated frequencies, $F_{av} = xF_{\rm P} + (1-x)F_{\rm As}$
where $F_{\rm P}/F_{\rm As}$ refers to the (average maximal and minimal extremal) frequencies calculated with the atom
set to P or As respectively. The calculated changes are much smaller than we observe experimentally, so it seems
unlikely that the shrinking electron sheets can be explained by conventional bandstructure theory.

An alternative is that the shrinking is driven by many-body effects. Recently, Ortenzi {\it et al.} \cite{Ortenzi09}
suggested that in LaFePO the shrinking of the electron and hole Fermi-surfaces and enhancement of effective masses can
be explained by the interaction between the electron and hole bands. The observed correlation of the Fermi surface
areas with the increase in effective mass and increase in $T_c$ would be consistent with this model; however, it
remains to be seen if quantitative agreement can be achieved.

In summary, we have presented dHvA data which shows the evolution of the Fermi surface of BaFe$_2$(As$_{1-x}$P$_x$)$_2$
as $x$ is varied towards $T_c^{\rm max}$. We find that the volume of the electron sheets (and via charge neutrality
also the hole sheets) shrink linearly and the effective masses become strongly enhanced with decreasing $x$. It seems
unlikely that these changes are a simple consequence of the one-electron bandstructure but instead they likely
originate from many-body interactions. These changes may be intimately related to the high $T_c$ unconventional
superconductivity in this system.

We thank J.\ Analytis for helpful discussions and E.A.\ Yelland for the use of computer code.  Part of this work has
been done with the financial support of the UK's EPSRC and Royal Society, EuroMagNET under the EU contract no.228043,
Scientific Research on Priority Areas of New Materials Science Using Regulated Nano Spaces (20045008), Grant-in-Aid for
Scientific Research on Innovative Areas ``Heavy Electrons (20102002), Specially Promoted Research (2001004) and
Grant-in-Aid for GCOE program `The Next Generation of Physics, Spun from Universality and Emergence' from MEXT, Japan.
Work performed at the NHMFL in Tallahassee, Florida, was supported by NSF Cooperative Agreement No. DMR-0654118, by the
State of Florida, and by the DOE.

\bibliography{BaFe2AsP2}
\bibliographystyle{aps5etal}

\end{document}